\documentclass[aps,prxquantum,floatfix,twocolumn,superscriptaddress]{revtex4-1}
\usepackage{latexsym}
\usepackage{graphicx}
\usepackage{rotating}
\usepackage[normalem]{ulem}
\usepackage{hyperref}
\usepackage{amsmath,amsfonts}
\usepackage{bm}
\usepackage{color}
\usepackage{textgreek}
\usepackage{soul}  
\usepackage{qcircuit}
\usepackage[greek,english]{babel}
\usepackage{amsmath}

\newcommand{\dn}{\downarrow}

\usepackage{color}

\usepackage{amsfonts}
\usepackage{mathrsfs}
\usepackage{physics}

\hypersetup{colorlinks=true,urlcolor=blue,linkcolor=blue}

\def\a{\alpha}

\def\b{\beta}
\def\D{\Delta}
\def\d{\delta}
\def\e{\epsilon}

\def\E{\mathcal{E}}
\def\h{\hat}

\def\s{\sigma}

\def\GS{\text{GS}}
\def\w{\omega}

\def\dd{\dagger}
\def\ua{\uparrow}
\def\da{\downarrow}

\providecommand{\d}[1]{\color{grey}[#1]\color{black}}

\begin{document}
\title{Quantum subspace expansion algorithm for Green's functions}
\begin{abstract}

\end{abstract}

\author{Fran\c{c}ois Jamet}
\email{francois.jamet@npl.co.uk}
\author{Abhishek Agarwal}
\author{Ivan Rungger}
\email{ivan.rungger@npl.co.uk}
\affiliation{National  Physical  Laboratory,  Teddington,  TW11  0LW,  United  Kingdom}

\begin{abstract}
We present an algorithm to compute Green’s functions on quantum computers for interacting electron systems, which is a challenging task on conventional computers. It uses a continued fraction representation based on the Lanczos method, where the wave functions are expanded as linear combination of basis states within a quantum subspace. While on conventional computers the cost of the computation grows exponentially with system size, limiting the method to small systems, by representing the basis states on a quantum computer one may overcome this exponential scaling barrier.
We propose a two-level multigrid Trotter time evolution for an efficient preparation of the basis states in a quantum circuit, which takes advantage of the robustness of the subspace expansion against Trotter errors. Using a quantum emulator, we demonstrate the algorithm for the Hubbard model on a Bethe lattice with infinite coordination, which we map to a 16 qubit Anderson impurity model within the dynamical mean field theory. Our algorithm computes the Green’s function accurately for both the metallic and Mott insulating regimes, with a circuit depth several orders of magnitude below what has been proposed using direct time evolution. The two-level multigrid time evolution reduces the number of Trotter steps required to compute the Green’s function to about four to six. We therefore expect that the method can be used on near term quantum computers for moderate system sizes, while allowing for scalability to larger circuit depths and qubit numbers on future fault tolerant quantum computers.
\end{abstract}
\maketitle

\section{Introduction}

The Green’s function (GF) is a central quantity in materials science simulations for computing observables of interest, such as the density of states (DOS), optical and electronic conductivities, or Raman spectra\cite{Economou, negf}. It is also at the core of many embedding methods \cite{negf,  book1}. On a conventional computer, the Green's function remains a challenging quantity to compute. This is particularly the case for strongly correlated materials, where standard methods such as density functional theory or the so called $GW$ approximation are insufficient, and one needs to use non perturbative methods such as the dynamical mean-field theory (DMFT)\cite{dmft,dr.ru.17}.
Recent advances in the development of quantum computers have led to increasing interest in quantum algorithms for obtaining quantities such as the Green's function on quantum devices. Several approaches have been used, targeting either near term noisy devices or future fault-tolerant quantum computers.

In Ref. \cite{troyer} the authors propose computing the GF in the real-time domain using the Trotter time evolution operator, and then performing a Fourier transform to obtain the GF in the frequency domain. However, to resolve the features of the GF on the real energy grid a large number of Trotter time steps is required, which puts this algorithm out of reach for near-term quantum computers. Moreover, the ground state preparation method used relies on quantum phase estimation, which also requires a very deep quantum circuit \cite{troyer,PhysRevX.8.041015,Berry2018}. Refs. \cite{Jaderberg2020,PhysRevResearch.2.033281,Keen2020} propose to reduce the circuit depth required for the time evolution by using a variational quantum algorithm (VQA), where the accuracy of the time evolution relies on the accuracy of the variational quantum circuit optimization.
A recent work proposes to use the McLachlan variational principle to compute the GF in real time \cite{arxiv.2203.12372,PhysRevResearch.2.033281}.

Other approaches \cite{2sites,mlde,Keen2020} operate directly in the frequency domain, and target the practical applicability on noisy intermediate scale quantum (NISQ) devices. Some of these methods are based on the Lehmann representation of the GF, which requires computing the excited states of the Hamiltonian. This can allow one to compute the Green's function in a restricted energy window, where only a subset of all excited states needs to be computed. However, it is challenging to regularize the GF in this energy window to take into account the omitted excited states. Alternatively, approximate forms of the GF, such as low energy parametrized expansions, have recently been proposed \cite{PhysRevB.105.115108}.
In a previous study \cite{kvqa}, we proposed the Krylov variational quantum algorithm (KVQA) to compute the GF, where one uses its continued fraction representation. The advantage of the KVQA is that, instead of obtaining the GF using the excited states, one uses the Krylov states, which allow for faster convergence of the GF also as the system size becomes large. To prepare the Krylov states one uses a parametrized quantum circuit within a VQA. The KVQA requires moderate circuit depths, making it applicable on NISQ devices whilst also allowing use on larger systems if the underlying VQA can be scaled.

While VQAs allow performing quantitative simulations on currently available NISQ devices, a number of questions remain open regarding their scalability to larger system sizes \cite{Cerezo2021,jules_rev}. Typically, VQAs require training the parameters of a quantum circuit to minimize a cost function, such as the total energy. One of the main challenges of VQAs is to prepare the target wave function accurately in a parametrized quantum circuit. To overcome this problem, one can, instead of directly preparing the full target wave function on a quantum computer (QC), represent the target wave function as linear combination of basis states, each of which can be prepared on a QC individually. This  approach has been proposed in a number of recent articles \cite{qsd_theory,Huggins2020, 2103.08563, McClean2017, Motta2020, 1909.08925,  Stair2020}, although using different names, such as quantum Krylov subspace expansion \cite{Stair2020}, quantum filter diagonalization \cite{1909.08925}, variational quantum phase estimation \cite{2103.08563}, and quantum subspace diagonalization (QSD) \cite{qsd_theory}. In this article we refer to it as quantum subspace expansion (QSE), since the approach is not restricted to diagonalization or variational tasks, and the basis space is not necessarily the Krylov space.
Within QSE the Hamiltonian matrix elements in the chosen basis are obtained on a quantum computer. The expansion parameters of the state are then obtained on a classical computer by solving a linear algebra problem.
While the involved quantum circuits are, in general, deeper than the ones used in traditional VQAs, they are still relatively short compared to the depths of circuits in algorithms such as quantum phase estimation (QPE). Thus, QSE algorithms have the potential of being run on near term quantum computers, and may offer improved scalability when compared to conventional VQAs.

In this article we extend the KVQA method to the QSE framework, where we compute the GF as continued fraction, and represent the Krylov states using the QSE. To prepare the basis states on a quantum computer we use the real time evolution. We propose a Trotter evolution with multiple time-step sizes, which allows reducing the required number of Trotter steps to construct the basis.
We first present our QSE algorithm for Green's functions (QSEG), and then describe how we construct the basis states with the two-level multigrid Trotter time evolution. We then demonstrate our algorithm for the Hubbard model on a Bethe lattice with infinite coordination using the DMFT.

\section{Quantum subspace expansion for the Green's function}
To obtain the Green's function one first needs to compute the ground state (GS). We therefore start by presenting the QSE method for the ground state calculation based on existing literature, and then present our QSE algorithm for the computation of the Green's function (QSEG).

\subsection{Ground state}
Within QSE, one expresses the states as linear combination of basis states. The QSE GS wave function of a given Hamiltonian operator $\hat H$ is written as
\begin{equation}
\ket{\GS} = \sum^{n_{\phi}}_{i=1} \phi^{\GS}_{i}\ket{\phi_i},
\label{eq:gsQSD}
\end{equation}
and fulfills the relation $\bra{\text{GS}}\h H\ket{\text{GS}} {=} E_{\text{GS}}$, where $E_\GS$ is the lowest expectation value of $\h H$ in the QSE basis.
The set $\lbrace\ket{\phi_i}\rbrace$ with $n_\phi$ elements forms the chosen basis, and $\phi_i^{\GS}$ are the complex-valued expansions coefficients. The closeness of the QSE ground state $\ket{\GS}$ to the true ground state can be systematically improved by increasing the space spanned by the basis set $\lbrace\ket{\phi_i}\rbrace$. We denote the difference between the true ground state energy of $\h H$ and the QSE GS energy $E_\mathrm{GS}$ as $\D E$. The number $n_\phi$ can be kept small by choosing basis functions based on knowledge of the properties of a desired target state, such as for example wave functions with a given number of electrons. To obtain the $\phi^{\GS}_i$ one computes the matrix elements of $\hat{H}$ in the chosen basis as
\begin{equation}
   H_{ij} = \bra{\phi_i}\hat{H}\ket{\phi_j},\label{eq:H_matrix}
\end{equation}
as well as the corresponding overlap matrix elements
\begin{equation}
   S_{ij} = \bra{\phi_i}\ket{\phi_j}, \label{eq:O_matrix}
\end{equation}
and solves the linear equation
\begin{equation}
  \bm{H} \bm{\phi}^{\GS} = E_{GS} \bm{S} \bm{\phi}^{\GS} \label{eq:schrodinger_qsd} \\
\end{equation}
for the lowest eigenvalue $E_{\GS}$ and its corresponding eigenvector $\bm{\phi}^{\GS}$.
As a matter of notation we use bold font symbols to refer to matrices and vectors.

In the QSE framework, a quantum device is used to compute all elements of $\bm{H}$ (Eq. (\ref{eq:H_matrix})) and $\bm{S}$ (Eq. (\ref{eq:O_matrix})), and then the linear system in Eq. (\ref{eq:schrodinger_qsd}) is solved on a conventional computer to obtain $\bm{\phi}^{GS}$ and $E_{GS}$.
The quantum circuits required to compute $\bm{H}$ and $\bm{S}$ strongly depend on the choice of the basis $\{\ket{\phi_i}\}$. We give the  details of the implementation as a quantum circuit in Appendix \ref{sec:HS}.
Since these matrices are Hermitian, there are $\frac{n_{\phi}(n_{\phi}+1)}{2}$ independent matrix elements that need to be computed for each matrix. Such linear systems of equations can be routinely solved on conventional computers for $n_\phi$ up to a few tens of thousands.
The construction of an efficient basis set, which allows to keep $n_\phi$ small, is of central importance for QSE to be efficient. We will discuss the basis set construction in Sec. \ref{sec:basis}.

The main advantage of QSE over the  variational quantum eigensolver (VQE) is that QSE does not include an optimization of the quantum circuit parameters as part of the minimization process, which can be an NP-hard problem \cite{vqe_nphard}. Furthermore, in VQE, each update of the quantum circuit parameters needs a new call to the quantum computer, whereas in QSE all required circuits can be sent to the quantum computer in just one call. An additional benefit of QSE is that it can be efficiently parallelized over many quantum computers, because each matrix element can be computed on a different quantum computer at the same time. Such a parallel implementation of QSE is key for its practical scalability as the basis set size increases. The drawback of QSE is that one does not have access to the full wave function directly on the quantum computer, since it is decomposed as linear combination of the basis functions.

\subsection{Green's function}
\label{sec:gf_method}
We now outline how Green's functions can be computed using a QSE based algorithm, which we denote as QSEG.
To obtain the GF, one needs to compute expectation values of the form
\begin{align}
    F_{\h A \h B}(z) = \bra{\Phi} \h A^\dd (z - \h H)^{-1} \h B\ket{\Phi},
    \label{eq:general_dyn}
\end{align}
where $\h A$ and $\h B$ can be arbitrary operators, $\ket{\Phi}$ is an arbitrary state, and $z$ is a complex- or real-valued energy. For the GF one considers the case when $\h B$ and $\h A$ are equal to the creation ($\h c^\dd_\a$) and annihilation ($\h c_\a$) operators of a particle in the orbital with integer index $\a$. The greater and lesser GFs are defined as
\begin{align}
    G^>_{\a\b}(z)& = \bra{\text{GS}}\hat{c}_{\a}(z - (\hat{H} -E_{\text{GS}}))^{-1}\hat{c}^\dd_{\b}\ket{\text{GS}}, \label{eq:greater} \\
    G^<_{\a\b}(z)& = \bra{\text{GS}}\hat{c}^\dd_{\a}(z + (\hat{H} -E_{\text{GS}}))^{-1}\hat{c}_{\b}\ket{\text{GS}}, \label{eq:lesser}
\end{align}
respectively. These can be used to obtain the retarded GF as
\begin{align}
    G_{\a\b}(z) &= G_{\a \b}^{>}(z) + G_{\a \b}^<(z).  \label{eq:retgGeneral}
\end{align}
The imaginary part of the retarded GF gives the DOS, with $\mathrm{DOS}(\omega)= -\frac{1}{\pi} \text{Im}G(\w + i\d)$, $\omega$ a real energy, and $\d$ a small positive number. To evaluate the quality of the GF for a given material one can compare the DOS to experimental measurements.

In the  rest of the description of the algorithm we  focus on the computation of the diagonal part of the  greater GF (Eq. (\ref{eq:greater})), so that $\beta=\alpha$. The algorithm for the lesser GF (Eq. (\ref{eq:lesser})) is analogous. The generalization to the off-diagonal elements is given in Appendix \ref{sec:offdiagonal}. In the rest of this article we drop the index $\a$ to simplify the notation. The extension to expectation values for arbitrary operators as given in Eq. (\ref{eq:general_dyn}) is analogous to the derivation below.

We use the continued fraction representation of the GF computed using the Lanczos scheme\cite{lanc1,lanc2,kvqa}. This is based on the construction of an orthogonalized Krylov basis set, $\{\ket{\chi_n}\}$, where the Hamiltonian is tridiagonal, so that the greater GF can be written as a continued fraction
  \begin{equation}
    G^>(z) = \frac{1}{z - a_0-\frac{b_1^2}{z - a_1-\frac{b_2^2}{z -  a_2....}}}. \label{eq:continued_fraction}
  \end{equation}
  The Krylov basis is constructed starting with $\ket{\chi_0} = c^\dd \ket{\GS}$, and then iterated to increasingly larger $n$ using the relations\cite{lanc1,lanc2,kvqa}
\begin{align}
  b_{n}^2 &=\expval{\hat{H}^2}{\chi_{n-1}}-a_{n-1}^2-b_{n-1}^2  ,\label{eq:bnp1}\\
  \ket{\chi_{n}} &= \frac{1}{b_{n}}[(\hat H-a_{n-1})\ket{\chi_{n-1}}-b_{n-1}\ket{\chi_{n-2}}], \label{eq:lanczos_iter} \\
  a_{n} &=\expval{\hat H}{\chi_{n}}.\label{eq:anp1}
\end{align}
Here $a_n,b_n$ are the coefficients used in Eq. (\ref{eq:continued_fraction}) ($b_0=0$). If all these orthogonalized Krylov states up to an integer $n$ are included in a basis set, then this spans the space corresponding to the non-orthogonal states $\ket{\chi_0}, \h H\ket{\chi_0},\h H^2\ket{\chi_0}\dots,\hat H^n\ket{\chi_0}$.
Importantly, in practical calculations one can stop the iterative process when one of the $b_n^2 {\approx} 0$, since in this case higher terms do not contribute significantly to the GF.

On conventional computers this procedure is used extensively, where it has been found to be an efficient way to compute the GF \cite{lanc1,lanc2,lanc_mps}. The main limitation is that the dimension of the Hilbert space scales exponentially with the size of the system, making this procedure impractical for large system sizes. On a QC, it may be possible to overcome this exponential scaling. In a previous study \cite{kvqa} we proposed a variational procedure to compute the $a_n$ and $b_n$ coefficients on a quantum computer. It overcomes the exponential scaling of the computation of matrix elements such as  $\expval{\hat H}{\chi_{n}}$, and relies on the optimization of the quantum circuit parameters via minimization of a cost function. Such optimization becomes challenging as the system size increases. The development of scalable VQE optimizers is an active area of research \cite{jules_rev}.

In this article we propose an alterntive approach based on the QSE framework. We project $\ket{\chi_n}$ on a subspace,  spanned by a basis set denoted as $\{ \ket{\psi_n}\}$, such that
\begin{equation}
  \ket{\chi_n} = \sum^{n_\psi}_{i=1} \psi^n_{i} \ket{\psi_{i}}.
  \label{eq:krylov_decomp}
\end{equation}
We note that the QSE $\ket{\chi_n}$ in this equation become exact if $\{\ket{\psi_{i}}\}$ is a complete basis of the Krylov space.
As noted in the previous section, the QSE basis has to be adapted to the type of states that it needs to represent. The basis set $\{ \ket{\psi_i}\}$ for the Krylov states in the GF is therefore different from the basis set for the GS $\lbrace\ket{\phi_i}\rbrace$ (Eq. (\ref{eq:gsQSD})). One major difference between the two basis sets is that while the GS basis set $\lbrace\ket{\phi_i}\rbrace$ only needs to be optimized to represent a single wavefunction (the ground state), the $\{ \ket{\psi_i}\}$ needs to be able to represent all the Krylov states.

The first Krylov state is determined by the relation $\ket{\chi_0} {=} c^{\dd}\ket{\GS}$, which combined with Eqs. (\ref{eq:gsQSD}) and (\ref{eq:krylov_decomp}) gives for the coefficients $\bm{\psi}^0$ of $\ket{\chi_0}$:
\begin{align}
    \bm{\psi}^0 & =   \bm{S}_{\psi}^{-1}\bm{S}_{\psi,\h c^\dd \phi}\bm{\phi}_{\text{GS}}.
\end{align}
Here, $\bm{\phi}_{\GS}$ is determined using the QSE method for the ground state presented in the previous section. The overlap, $\bm{S}_{\psi}$, and transition matrices, $\bm{S}_{\psi,\h c^\dd \phi}$, are given by
\begin{align}
     ( \bm{S}_{\psi,c^\dd \phi})_{ij}& = \bra{\psi_i} \h c^\dd\ket{\phi_j},\label{eq:O_matrix_phi0}
     \\ (\bm{S}_{\psi})_{ij}&  = \bra{\psi_i}\ket{\psi_j}. \label{eq:O_matrix_phi0C}
\end{align}
Then, to compute the vector $\bm{\psi}^{n}$, we insert Eq. (\ref{eq:krylov_decomp}) into Eqs. (\ref{eq:bnp1}-\ref{eq:anp1}), which gives
\begin{align}
  b^2_n &=  \bm{\psi}^{n-1 \dd }\bm{H}_{\psi}  \bm{S}^{-1}_{\psi}\bm{H}_{\psi}\bm{\psi}^{n-1}    -a_{n-1}^2 -b^2_{n-1} ,\label{eq:bnc} \\
  \bm{\psi}^{n} &=  \frac{1}{b_n}((\bm{S}_{\psi}^{-1 }  \bm{H}_{\psi}- a_{n-1}) \bm{\psi}^{n-1} - b_{n-1}\bm{\psi}^{n-2} )), \label{eq:psinc}\\
  a_n &=  \bm{\psi}^{n \dd } \bm{H}_{\psi} \bm{\psi}^{n},\label{eq:anc}
\end{align}
where the Hamiltonian matrix projected on the $\lbrace\ket{\psi_i}\rbrace$ basis set, $\bm{H}_\psi$, is given by
\begin{equation}
(\bm{H}_{\psi})_{ij} = \bra{\psi_i} H\ket{\psi_j} .\label{eq:H_matrix_g}
\end{equation}
In analogy to the GS computation, the matrix elements in Eqs. (\ref{eq:O_matrix_phi0}), (\ref{eq:O_matrix_phi0C})  and (\ref{eq:H_matrix_g}) are computed using a quantum device, while the iterative Eqs. (\ref{eq:bnc}-\ref{eq:anc}) are evaluated on a classical computer.
The computed $a_n$ and $b_n$ coefficients are then inserted in Eq.  (\ref{eq:continued_fraction}) to obtain the greater GF.

\section{Construction of the basis}
\label{sec:basis}

The procedure described in the previous section is valid for any choice of the bases $\{\ket{\phi_i}\}$ and $\{\ket{\psi_i}\}$. The accuracy of the GF determined by the QSEG algorithm relies on the construction of bases that efficiently describe the Krylov states $\ket{\chi_n}$.
In this section we present a systematic approach to  construct $\ket{\chi_n}$ for the QSEG method. We first outline how to construct the basis for the ground state based on existing literature using a Trotter time evolution. We introduce a two-level multrigrid time evolution grid to reduce the required number of Trotter steps. We then present the basis for the Green's function.

\subsection{Ground state basis}
\label{sec:gsbasis}

Several possibilities have been proposed in the literature for the QSE ground state basis, such as the use of a variational ansatz \cite{Huggins2020}, the use of operators contained in the Hamiltonian \cite{Lim2021}, imaginary time evolution \cite{Motta2020} and real time evolution \cite{cortes2021quantum,Stair2020,1909.08925}. In this paper we construct the basis with the real-time evolution operator \cite{2103.08563,qsd_theory,Stair2020}, which is a general and systematic approach. This approach allows to construct a basis that spans the space of the Krylov states built on the initial $\ket{\phi_0}$. If $\ket{\phi_0}$ has finite overlap with the ground state, then this basis gives exponentially fast convergence to the correct ground state of $\h H$ with respect to the number of basis states \cite{GoluVanl96}.

The basis for the GS (Eq. (\ref{eq:gsQSD})) is constructed by a repeated application of the time evolution operator $V(t) {=}e^{-it\h H}$ on a reference state $\ket{\phi_0}$, using time steps $\Delta_{t}$:
\begin{equation}
  \ket{\phi_l}  = (\h V(\D_t))^l \ket{\phi_0}. \label{eq:basis_normal}
\end{equation}
The reference state $\ket{\phi_0}$ is usually chosen in such a way that it is easy to prepare on a quantum computer, and that it also has a significant overlap with the GS. The chosen GS reference state depends on the Hamiltonian; in case of interacting electron systems, the Hartree Fock solution is a natural choice \cite{Stair2020,2103.08563}. Another possibility is to use an approximate VQE solution \cite{PhysRevC.105.024324}. For the integer indices $l$ we use the range from $-n_l$ to $n_l$. Such a positive and negative time evolution is proposed in Ref. \cite{1909.08925}, with $(\h V(\D_t))^{-l} {=}(\h V^{\dd}(\D_t))^l$. For a given quantum circuit that represents $V(\D_t)$, the circuit for the conjugate transpose operation can be obtained by changing the order of gates and signs of rotations appropriately.

The size of the basis is given by $n_{\phi}{=}2n_l+1$, and the circuit depth to implement $(\h V(\D_t))^l$ is equal to $d_V l$, where $d_V$ is the depth of the quantum circuit to  implement the time evolution operator $\h V(\D_t)$ performed using a Suzuki-Trotter expansion. Since the noise induced by 2-qubit gates is significantly larger than the one induced by single qubit gates, we equate the circuit depth to be the number of layers of 2-qubit gates. Within one layer, we assume that the gates can be executed in parallel.
As outlined in Ref. \cite{1909.08925}, if  $\D_t$  is too small, the basis vectors for $l$ and $l{+}1$ are very similar, so that the convergence of the QSE with respect to increasing $n_\phi$ is slow. On the other hand, if $\D_t$ is too large, the QSE might not converge to accurate energies, since high-frequency features obtained using small $\D_t$ are also required. Therefore, for each system there is an optimal $\D_t$ and $n_\phi$.

\begin{figure}
  \centering
  \includegraphics[scale=0.7]{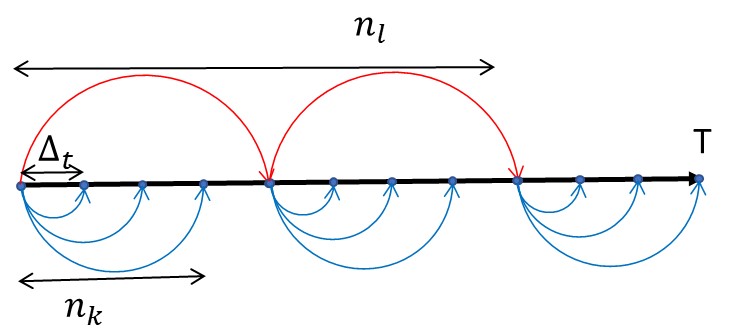}
  \caption{Schematic representation of the two-level multigrid Trotter time evolution basis. To construct a single basis state one first applies a number of large Trotter steps of length $(n_k{+}1)\D_t$ (indicated in red), followed by a single smaller step (indicated in blue).}
  \label{fig:schematic}
\end{figure}
For a fixed $\D_t$, $n_\phi$ needs to be increased sufficiently in order to resolve long time features, which requires longer quantum circuits. This brings it out of reach of near-term quantum hardware, which have moderately long coherence times and limited gate fidelities. To reduce circuit depths one may use approximate approaches for the time evolution, such as the ones based on the McLachlan variational principle \cite{Yuan2019}, or on variational fast forwarding \cite{Crstoiu2020}. Here we propose an alternative approach to reduce the required number of Trotter steps and with it the circuit depth, which is based on a multigrid time evolution, where time steps with varying sizes are used. We define the two-level multigrid time evolution basis as
\begin{equation}
\ket{\phi_{lk}}   =   \h V(k\D_t)(\h V((n_k+1)\D_t))^l  \ket{\phi_{0}},  \label{eq:basis}
\end{equation}
so that the GS wavefunction of Eq. (\ref{eq:gsQSD}) becomes
\begin{equation}
\ket{\mathrm{GS}} {=} \sum^{n_l}_{l=-n_l} \sum^{k_\mathrm{max}}_{k=k_\mathrm{min}} \phi^\mathrm{GS}_{lk} \ket{\phi_{lk}}.
\label{eq:GS2t}
\end{equation}
For $l>0$ we have $k_\mathrm{min}=0$ and $k_\mathrm{max}=n_k$, for $l<0$ we have $k_\mathrm{min}=-n_k$ and $k_\mathrm{max}=0$, and for $l=0$ we have $k_\mathrm{min}=-n_k$ and $k_\mathrm{max}=n_k$. This basis is constructed by a combination of large time-steps equal to $(n_k+1)\Delta_t$, and of smaller time steps between these large time steps, with step size that goes from $\Delta_t$ to $n_k\Delta_t$. The two-level multigrid is schematically illustrated in Fig. \ref{fig:schematic}. It allows to reach large total times $T$, while keeping a fine resolution of small time steps, with a reduced circuit depth when compared to the use of a single time-step size.

The number of basis states is equal to $n_{\phi} {=} 2(n_l{+}1)(n_k{+}1){-}1$, and the  circuit depth to implement $\h V(k\D_t)(\h V((n_k+1)\D_t))^l$ is equal to $d_V(1+l)$.  Therefore, for a fixed $\Delta_t$, increasing $n_k$ increases the maximum time reached, $T{=}((n_k+1)n_l+n_k)\D_t$, without increasing the maximum circuit depth, which is given by $d_V (n_l+1)$. The price to pay for increasing the time step is the increase of the Trotter error. However, QSE using real-time evolution is expected to be relatively tolerant to Trotter errors, because it only needs to add new degrees of freedom to the basis, so that the time-evolution itself does not need to be very accurate \cite{1909.08925}. Our multigrid approach exploits this by using a combination of large and small Trotter steps to reach long times with a fine resolution. In Sec. \ref{sec:results} we evaluate the performance of this multigrid time evolution approach for QSEG.

\subsection{Basis for the Green's function}

We now move to the construction of the basis to build the Krylov states for the GF (Eq. (\ref{eq:krylov_decomp})). The orthogonalized Krylov states are defined iteratively in Eqs. (\ref{eq:bnp1}-\ref{eq:anp1}). As discussed there, these states are linear combinations of powers of $\h H$ applied to the state $\ket{\chi_0}{=}c^\dd \ket{\GS}$.
To construct the basis we therefore use a real-time evolution, in analogy to the construction of the basis for the GS, but with the initial state $\ket{\chi_0}$. The reason for this choice is that as one combines states for increasing number of Trotter time-steps applied to $\ket{\chi_0}$, one covers the space spanned by powers of $\h H$ applied to $\ket{\chi_0}$.

We set the first basis state as $\ket{\psi_0}{=}c^\dd \ket{\GS}$, so that $\ket{\chi_0}{=}\ket{\psi_0}$.
Using the two-level multigrid time evolution we then obtain the other basis states as
\begin{equation}
\ket{\psi_{lk}}   =   \h V(k\tilde{\D}_t)(\h V((\tilde{n}_{k}+1)\tilde{\D}_t))^l  \ket{\psi_{0}},\label{eq:chiQSD}
\end{equation}
where $\tilde{\D}_t$ is the step size. The  circuit depth to implement $\h V(k\tilde{\D}_t)(\h V((\tilde{n}_k+1)\tilde{\D}_t))^l$ is equal to $d_V(1+l)$. The Krylov states in Eq. (\ref{eq:krylov_decomp}) become
\begin{equation}
\ket{\chi_n} {=} \sum^{\tilde{n}_l}_{l=-\tilde{n}_l} \sum^{\tilde{k}_\mathrm{max}}_{k=\tilde{k}_\mathrm{min}} \psi^n_{lk} \ket{\psi_{lk}}.
\label{eq:chin2t}
\end{equation}
For $l>0$ we have $\tilde{k}_\mathrm{min}=0$ and $\tilde{k}_\mathrm{max}=\tilde{n}_k$, for $l<0$ we have $\tilde{k}_\mathrm{min}=-\tilde{n}_k$ and $\tilde{k}_\mathrm{max}=0$, and for $l=0$ we have $\tilde{k}_\mathrm{min}=-\tilde{n}_k$ and $\tilde{k}_\mathrm{max}=\tilde{n}_k$. We use the notation with a tilde to refer to parameters of the basis set $\lbrace\ket{\psi_{lk}}\rbrace$ for the Krylov states to distinguish them from the parameters of $\lbrace\ket{\phi_{lk}}\rbrace$. The basis size for the Krylov states is $n_{\psi} {=} 2(\tilde{n}_l{+}1)(\tilde{n}_k{+}1){-}1$, the maximum circuit depth is $d_V(1+\tilde{n}_l)$.

To obtain $\ket{\chi_0}$ one can use a number of approaches. Using a variational quantum algorithm (VQA) one can directly obtain $\ket{\chi_0}$ as the output of a quantum circuit \cite{kvqa}. Alternatively, one can represent the GS directly on a quantum computer, for example by using quantum phase estimation \cite{troyer} or the Rodeo algorithm \cite{rodeo} if the quantum device allows for a deeper circuit. Since the creation operator $c^\dd$ can typically be expressed as a sum of two Pauli strings on a quantum computer, one can then obtain $\ket{\chi_0}$ as a sum of two terms. This can be generalized if one uses QSE for the GS, where $\ket{\chi_0}$ can be expressed as a linear combination of the chosen basis states.

Here we choose to use QSE also to compute the ground state. We note that QSEG is independent of whether QSE is also used for the construction of $\ket{\chi_0}$, since QSEG can be used with any of the aforementioned approaches to either obtain $\ket{\chi_0}$ via the GS, or to obtain $\ket{\chi_0}$ directly.
With the QSE GS wave function in Eq. (\ref{eq:GS2t}) we obtain
\begin{equation}
\ket{\psi_{0}} {=} \sum^{n_l}_{l=-n_l} \sum^{k_\mathrm{max}}_{k=k_\mathrm{min}} \phi^\mathrm{GS}_{lk} c^\dd\ket{\phi_{lk}}.
\label{eq:psi0}
\end{equation}
The total maximum number of Trotter steps to construct the basis set $\lbrace\ket{\psi_{lk}}\rbrace$ from $\ket{\phi_0}$ then is $(n_l+\tilde{n}_l+2)$, for any $n_k{>}0$ and $\tilde{n}_k{>}0$. If $n_k=0$ and $\tilde{n}_k=0$, then the method corresponds to a normal single-step Trotter time evolution.

In this basis, the Hamiltonian matrix elements in Eq. (\ref{eq:H_matrix_g}) become
 \begin{align}
  \bra{\psi_{k'l'}}\h H\ket{\psi_{kl}} {=}&\nonumber\\
  \sum_{l_1k_1l_2k_2}  (\phi_{k_2l_2}^\GS)^*  \phi_{k_1l_1}^\GS & \bra{\phi_{0}}  \h W^\dd_{l'k'l_2k_2}\h H  \h W_{lkl_1k_1} \ket{\phi_{0}},
  \label{eq:hpsielement}
\end{align}
where  $\h W_{lkl_1k_1} {=} \h V(k\tilde{\D}_t)(\h V((\tilde{n}_{k}{+}1)\tilde{\D}_t))^l c^\dd  \h V(k_1\D_t)$ $ (\h V((n_k{+}1)\D_t))^{l_1}$ is used to simplify the notation.  The method to compute the overlap matrices in Eqs. (\ref{eq:O_matrix_phi0}) and (\ref{eq:O_matrix_phi0C}) is analogous.
Equation (\ref{eq:hpsielement}) shows that each of the $n_{\psi}^2$ elements of $\bm{H}_\psi$ is a sum of $n_{\phi}^2$ matrix elements. Therefore, a total of $\frac{(n_{\psi}+1)n_{\psi}}{2}n^2_{\phi}$ matrix elements need to be computed on a quantum device. We give the quantum circuit implementation to compute these matrix elements in Appendix \ref{sec:HS}, where we also show that the required circuit depth to evaluate one matrix element in Eq. (\ref{eq:hpsielement}) is $2(d_0+(l+\tilde{l}+2)d_V)$. Here $d_0$ is the depth required to construct the state $\ket{\phi_0} {+} \ket{0}$ needed for the multi-fidelity protocol outlined in Appendix \ref{sec:HS}. The maximum required circuit depth for the QSEG method therefore is
\begin{equation}
  D_\mathrm{max} = 2(d_0+(n_l+\tilde{n_l}+2)d_V).
  \label{eq:depth}
\end{equation}

The number of matrix elements can become large for increasing basis set size. To reduce the number of elements one can aim to minimize $n_\phi$ for a given target accuracy, or also implement the GS or $\ket{\phi_0}$ directly on a quantum device with the methods outlined above. Importantly, these matrix elements can be computed independently, so that the runtime can be efficiently decreased by parallel runs over multiple sets of qubits or quantum computers.

\section{Application to the Anderson impurity model}
\label{sec:aim}
To demonstrate the QSEG algorithm on a particular system we apply it to the Anderson impurity model (AIM). The AIM consists of a set of $n_\mathrm{imp}$ interacting impurities embedded in a set of $n_\mathrm{b}$ bath sites. On the impurity sites electron-electron interactions are explicitly included in the Hamiltonian, while these are absent in the bath sites. The multi-orbital AIM Hamiltonian with density-density interaction is given by
\begin{align}
  \h H &= \h H_0 + \h H_{\text{int}}, \\
 \h H_0 &= \sum^{n_b+n_{\text{imp}} }_{i,j=1} \sum_{\s \in \{\ua,\dn\}}\e_{ij} \h c^\dd_{i\s}\h c_{j\s}, \\
 \h H_\mathrm{int} &= \sum_{i,j=1}^{n_\text{imp}}\sum_{\s,\s' \in \{\ua,\dn\}}U^{\s\s'}_{ij} \h n_{i\s} \h n_{j \s'}, \label{eq:aim}
\end{align}
where $\h c_{i\s}^\dd$ and $\h c_{i\s}$ are the creation and annihilation operators of an electron of spin $\s$ on site $i$, respectively. The site indices are ordered in such a way that the first $n_\mathrm{imp}$ sites are the impurities, while the subsequent $n_\mathrm{b}$ sites are the bath, for a total of $N{=}n_{\text{imp}}+n_b$ sites. $\hat{H}_0$ is the non-interacting part of the Hamiltonian, which acts on all the sites, and $\e_{ij}$ are the onsite energy and hopping matrix elements. $H_\mathrm{int}$ is the local Coulomb interaction acting on the impurities, with $\h n_{i\s} {=} \hat{c}^\dd_{i\s}\hat{c}_{i\s}$, and $U_{ij}^{\s\s'}$ is the Coulomb interaction magnitude.
We use the Jordan-Wigner transform \cite{jordan1928paulische} to map this fermionic Hamiltonian to a bosonic spin Hamiltonian. Each spin-orbital $(i,\s)$ is mapped to a bosonic orbital such that $\h  c^\dd_\a {=} (\prod_{\b =1}^{\a- 1}\h \s^{\b}_Z)(\h \s^\a_X - i \h \s^\a_Y)$, with $\a{=} i$ if $\s {=}\ua$ and $\a{=} N+i$ if $\s {=}\da$.

\begin{figure}
   \begin{minipage}[c]{.24\textwidth}
     \[ \Qcircuit @C=1em @R=1em {
       &\gate{H}  &\ctrl{1} &\gate{R_y(\frac{\theta}{2})} &\ctrl{1} &\gate{H} \\
       &\qw  &\targ & \gate{R_y(\frac{\theta}{2}) }& \targ & \qw \\} \]
     \end{minipage}\hfill
     \begin{minipage}[c]{.24\textwidth}
         \[= \begin{pmatrix}
             1 & 0 &0&0 \\
             0 &  \cos(\frac{\theta}{2}) & \sin(\frac{\theta}{2}) & 0\\
             0 & -\sin(\frac{\theta}{2}) & \cos(\frac{\theta}{2}) & 0\\
             0 & 0 & 0 & 1 \\
           \end{pmatrix}\]
       \end{minipage}
 \caption{Diagram of the quantum circuit  to implement the Givens rotation required for the operator $\hat {C}$ in Eq. \ref{eq:H0}}
 \label{fig:givens}
 \end{figure}
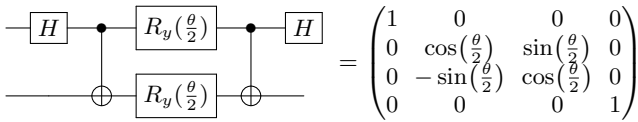

Since $\h H_0$ is a quadratic Hamiltonian, the quantum circuit for the time evolution $e^{-it \h H_0}$ can be implemented exactly using Givens rotations \cite{PhysRevLett.120.110501}. $\h H_\mathrm{int}$ is a sum of commuting terms, so that the expansion $e^{-i\Delta_t \h H_\mathrm{int}}{=}\prod_k e^{-i\h h_k\Delta_t}$ is exact, where $h_k$ are the commuting terms in $\h H_\mathrm{int}$.
$H_{int}$ and $H_0$ do not communte. The time evolution $e^{-iH \D_t}$ can be approximatively perfomed using  the  Suzuki-Trotter expansion of the time evolution operator $ \hat{V}(\Delta_t)=e^{-i\Delta_t \h H}=e^{-i\Delta_t  \left(\h H_0+\h H_\mathrm{int}\right) }$ in terms of $e^{-i\Delta_t \h H_\mathrm{int}}$ and  $e^{-i\Delta_t \h H_\mathrm{0}}$. To reduce the Trotter-error to $O(\Delta_t^3)$ we use the symmetrized form of the expansion \cite{Kivlichan2020improvedfault}, resulting in
\begin{equation}
  \hat{V}(\Delta_t) \approx e^{-i \frac{\Delta_t}{2} \h H_\mathrm{int}} e^{-i \Delta_t \h H_{0}} e^{-i \frac{\Delta_t}{2} \h H_\mathrm{int}}.
  \label{eq:time_evolution}
\end{equation}
Since $\h H_0$ is a quadratic Hamiltonian, its time evolution up to a global phase can be implemented exactly in a quantum circuit as \cite{Wecker2015}
\begin{equation}
  e^{-i \D_t \hat{H}_0} = \hat{C}^\dd \prod_{j=1}^{N} e^{-i \D_t \frac{\E_j}{2} (1-\hat{\s}^j_Z)} \;e^{-i \D_t \frac{\E_j}{2} (1-\hat{\s}^{j+N}_Z)
  } \hat{C},
  \label{eq:H0}
\end{equation}
where $\E_j$ are the  eigenvalues of the matrix $\e_{ij}$ and $\h C$ is a unitary transformation rotating the local basis to the orbital basis where $H_0$ is diagonal. As illustrated in Ref. \cite{PhysRevLett.120.110501}, $\h C$ can be implemented with $N$ layers of Givens rotations, each requiring 2 controlled-NOT (CNOT) gates, as shown in Fig. \ref{fig:givens}. Therefore, the total number of CNOT layers for applying $e^{-i\hat{H}_0 \D_t}$ in Eq. (\ref{eq:H0}) results in $4N$ layers of CNOT gates in the quantum circuit.

The depth of $e^{-iH_\mathrm{int}\D_t}$ depends of the number of impurities and on the type of interactions. In the case of the AIM with a single impurity, the interaction, once the Jordan-Wigner is performed, is given by $H_{int} {=} \frac{U}{4}( \s_Z^1 \s_Z^N-\s_Z^1 - \s_Z^N  )$. All these terms commute with each other, and hence its exponentiation can be decomposed into three separate terms. Only the exponentiation of the $\s_Z^1 \s_Z^N$ term requires CNOT gates, namely two. Combining it with the Givens rotations for $H_0$ we therefore have $d_V {=} 4N{+}4$ for the single-impurity AIM. Note that the method can also be ported to other interaction terms, such as the Hubbard-Kanomori interactions, where an approach using compressed double-factorized Hamiltonians can be considered \cite{Cohn_2021}.

As reference state $\ket{\phi_0}$ we use the minimal energy state of $\hat{H}_0$, which has $n_\ua$ spin-up and $n_\da$ spin-down particles.
To construct such state, we first apply a circuit to fill the particles in the orbital basis, where $\hat{H}_0$ is diagonal, and then rotate the state to the local basis using $\hat{C}^\dd$
\begin{equation}
  \ket{\phi_0} = \hat{C}^\dd \prod_{i=1}^{n_\ua} \s_X^i\prod_{j=N}^{n_\da+N} \s_X^j \ket{0}.
  \label{eq:phi_0}
\end{equation}
This state can be implemented with $2N$ layers of CNOT gates. To use the multi-fidelity estimation protocol outlined in Appendix \ref{sec:HS}, we need to construct the state $\ket{\phi_0} + \ket{0}\otimes \ket{0}...\otimes \ket{0}$, which requires an addition of $\mathrm{max}(n_\ua,n_\da){-1}$ CNOT layers, as shown in Appendix \ref{sec:HF_prepa}. The total circuit depth to construct these states is therefore given by $d_{0}{=}2N+\text{max}(n_\ua,n_\da)-1$.

 \section{Results}
\label{sec:results}

We apply the QSEG method to the prototypal single-band Hubbard model, the Bethe lattice \cite{dmft} with infinite coordination at half filling. The GF is obtained by the DMFT, which maps the Hubbard model to an AIM, where the bath hybridization $\D(z)$ is determined as $\D(z) {=} \gamma^2G(z)$. Here $\gamma$ is the Bethe lattice hopping amplitude and $G(z)$ the lattice local retarded GF. Since $G(z)$ itself depends on $\D(z)$, the DMFT equation needs to be solved self-consistently. In the rest of this section we work in units where $\gamma=1$. Using these units the AIM GF at DMFT self-consistency is equal to the bath hybridization ($\D{=}G[\D]$). The Coulomb interaction magnitude of the AIM is equal to the one of the Hubbard model.

\begin{figure}
    \centering
    \includegraphics[scale=0.3]{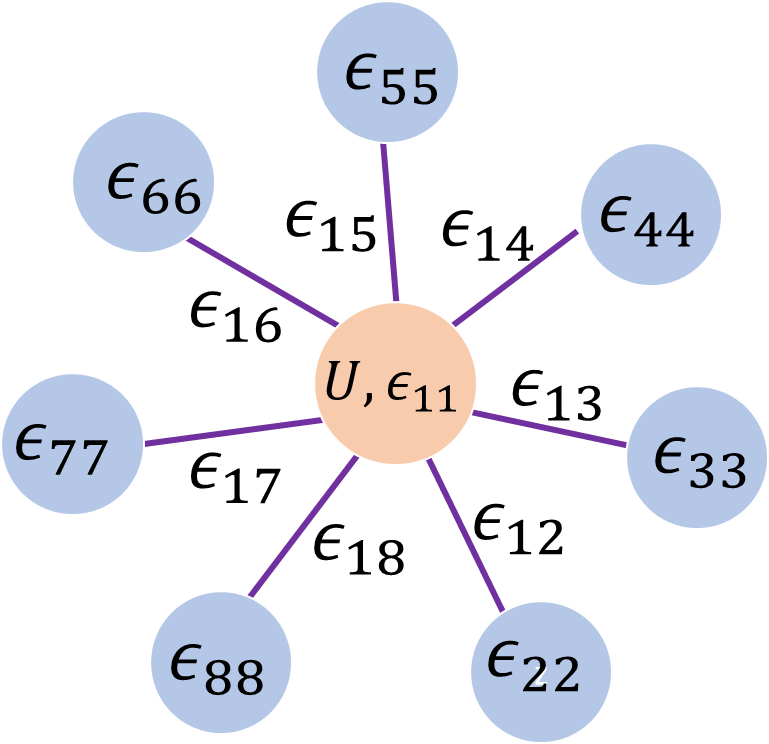}
    \caption{Schematic representation of the Anderson impurity model (AIM) with 1 impurity (orange circle) and 7 bath sites (blue circles), indicating all the AIM parameters $\epsilon_{ij}$ and $U$.}
    \label{fig:aim}
\end{figure}
To map the system to a finite number of qubits we discretize the bath to a finite number of bath sites, a method known as exact diagonalization \cite{dmft}, with DMFT self-consistency parameters  $\e_{ij}$ in Eq. (\ref{eq:aim}).  This model has been addressed on quantum hardware with 1 site in the bath \cite{2sites,Keen2020,Jaderberg2020}, and on quantum computing emulators with 5 sites in the bath \cite{troyer}.
In this article we use 7 sites in the bath. In total we have 1 impurity + 7 bath sites, so that we require 16 qubits. We use the quantum emulator Qulacs \cite{qulacs} to run all the simulations. We focus on the half-filling case, where $\e_{11}=-\frac{U}{2}$ in Eq. (\ref{eq:aim}) ($n_\mathrm{imp}=1$).
We start the DMFT self-consistency loop from the non-interacting solution, and iterate to convergence.
Further details of the DMFT self-consistent procedure are given in Appendix \ref{sec:dmft}.

We first determine the convergence behavior of the ground state, and then evaluate the quality of the GF obtained using the QSEG method. We present the results for both the metallic and the insulating regime of the model.

\subsection{Ground state}

To obtain the GF one first needs to compute the GS wave function and energy. A key element of the QSE basis for the ground state is the reference state $\ket{\phi_0}$. If the reference state is closer to the ground state, the convergence of $\D E$ with respect to the size of the basis is  generally improved.
In the case of the AIM, we choose  to take the solution of the AIM for $U{=}0$ as reference state $\ket{\phi_0}$.
This corresponds to the Hartree Fock solution, which can be constructed as a  Slater determinant and implemented efficiently on a quantum computer \cite{Wecker2015,PhysRevLett.120.110501}.

\begin{figure}
  \centering
  \includegraphics[scale=0.5]{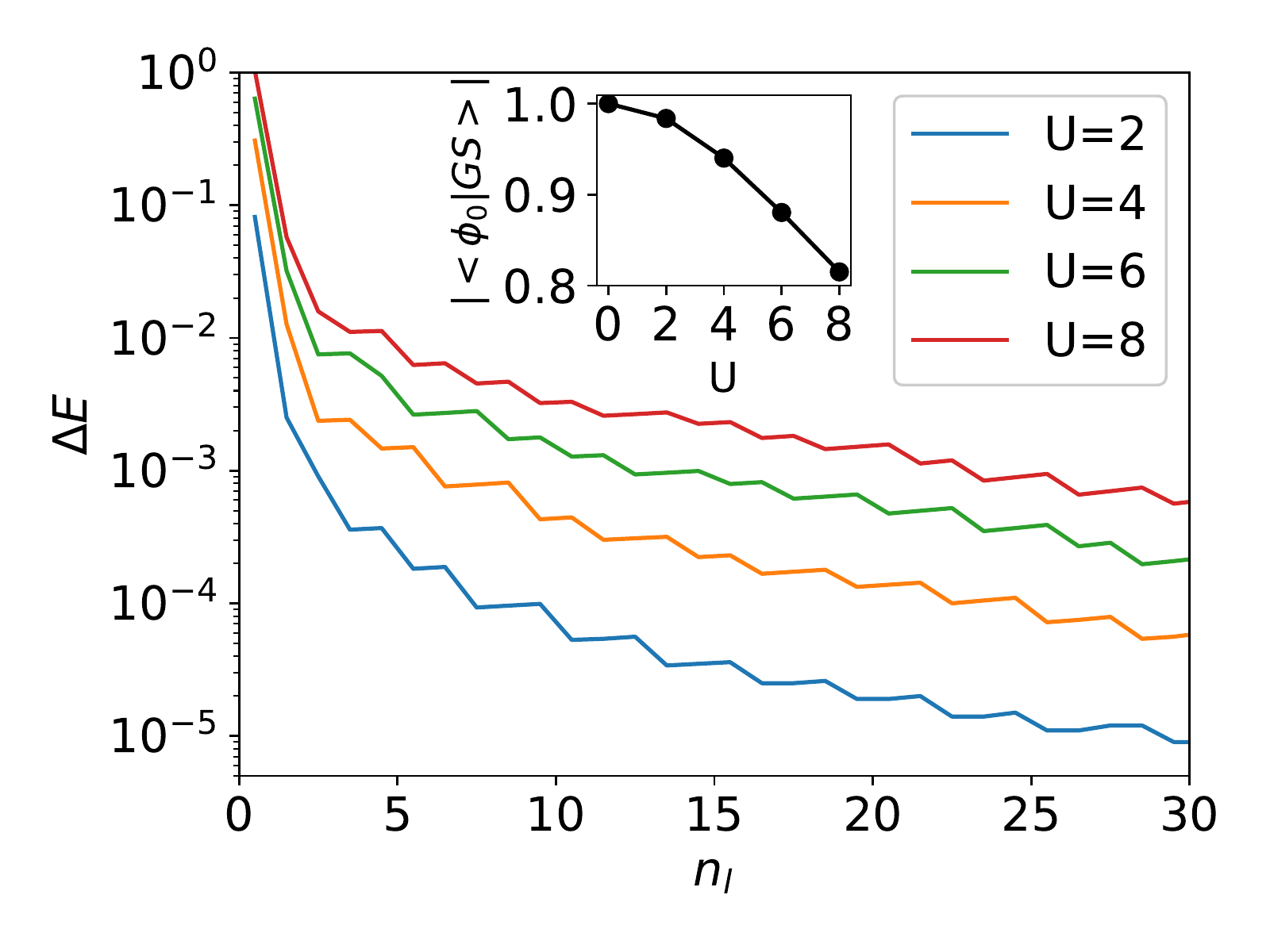}
  \caption{Comparison of the error in the ground state energy $\Delta{E}$ obtained with the QSE algorithm for different Coulomb interaction strengths $U$, as a function of the number of Trotter steps $n_l$ used to construct the QSE GS basis. The AIM parameters are the ones for the first step in the self-consistent DMFT cycle (the values are given in Appendix \ref{sec:dmft}) ($\D_t{=}0.1$, $n_k{=}0$). The inset shows that when $U$ increases, the overlap between $\ket{\phi_0}$ and the ground state decreases, resulting in a slower reduction of $\Delta E$ with increasing $n_l$.}
  \label{fig:phi0}
\end{figure}
We first evaluate how well the QSE energy converges towards the true GS energy for increasing number of Trotter steps. Here we use the AIM parameters of the first DMFT iteration, which corresponds to the non-interacting hybridization function. The values of $\epsilon_{ij}$ for this case are given in Appendix \ref{sec:dmft}.
Fig. \ref{fig:phi0} shows the evolution of the error of the QSE GS energy $\Delta{E}$ with respect to increasing $n_l$ for different values of $U$, with $\D_t{=}0.1$ and $n_k{=}0$. It shows that by increasing $n_l$ one can systematically reduce $\Delta{E}$. For all values of $U$ the error goes below $10^{-2}$ already at small $n_l$ values, after which the rate of decrease is slower, in particular for larger $U$.
For large $n_l=30$ the error is small, of the order of $10^{-5}$ to $10^{-3}$ for different values of $U$. The required accuracy in $\Delta{E}$, and with it the required $n_l$, depends on the needed accuracy on the Green's function and its energy resolution.
The reason that for a given $n_l$ the error increases with $U$ is that the reference state $\ket{\phi_0}$ has progressively less overlap with the ground state, as shown in the inset of  Fig. \ref{fig:phi0}. This is due to our choice of reference state being equal to the GS at $U=0$. For larger $U$ it may eventually be advantageous to use different reference states, for example, the state corresponding to the $U \rightarrow \infty$ limit.

\begin{figure}
  \centering
  \includegraphics[scale=0.4]{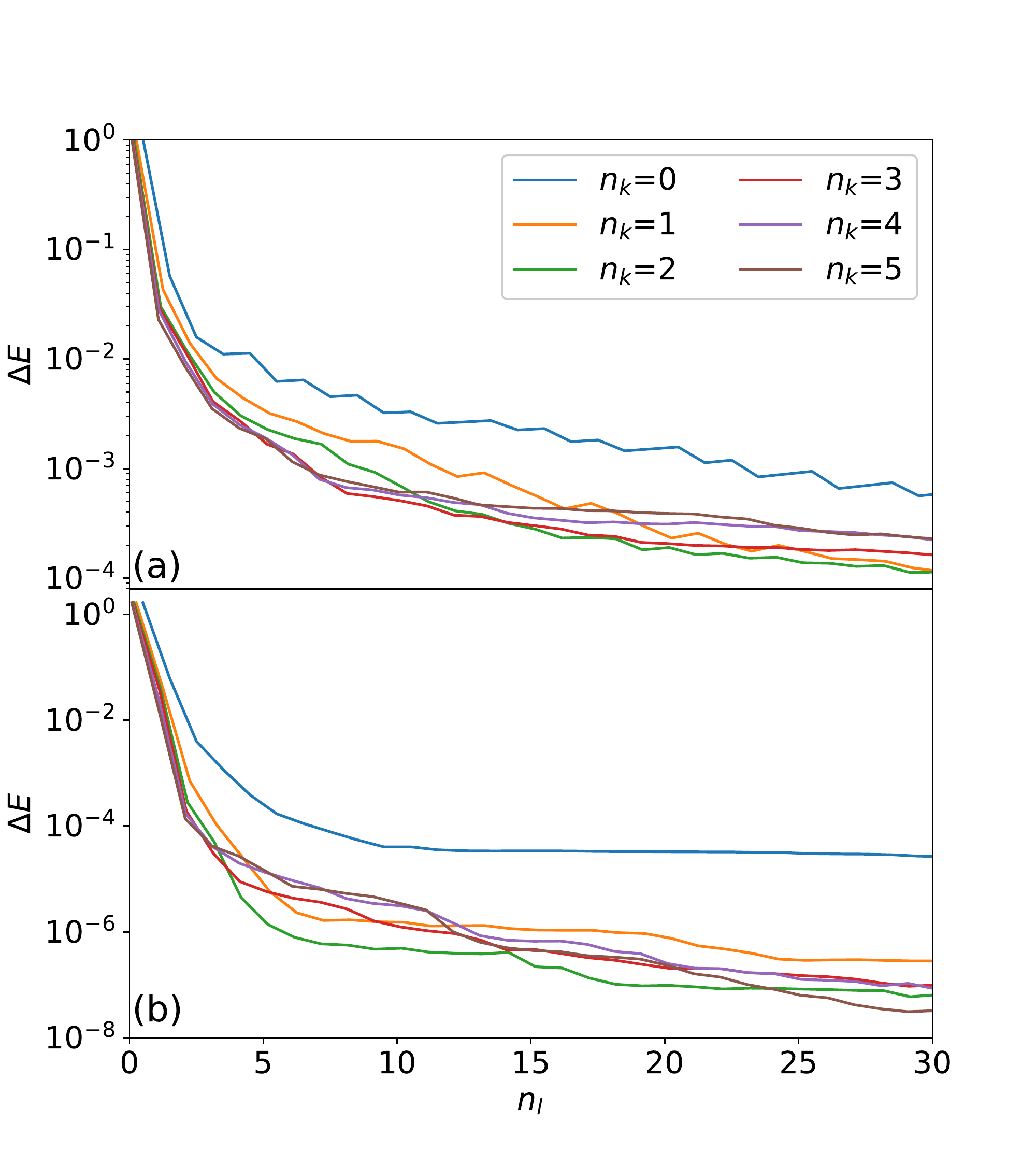}
  \caption{Analogous plots to Fig. \ref{fig:phi0}, replacing the single-step Trotter evolution ($n_k=0$) with the two-level multigrid evolution ($n_k>0$). In (a) the AIM parameters are the ones for the first step in the self-consistent DMFT cycle (identical to the ones used in Fig. \ref{fig:phi0}), while in (b) they are the ones at DMFT self-consistency (values are given in Appendix \ref{sec:dmft}). We note that the (a) and (b) plots have different scales for the vertical axis ($\D_t{=}0.1$). The two-level multigrid time evolution reduces the error significantly faster than the single-step time evolution.}
  \label{fig:egs_kmax}
\end{figure}
For the rest of the study on the GS energy, we focus on the case $U{=}8$, since it is the most challenging. Since the maximum circuit depth is a major constraint on near term QCs, we evaluate if the two-level multigrid time evolution basis introduced in Sec. \ref{sec:gsbasis} allows reducing the number of Trotter steps, and with it the total circuit depth required for a given target $\Delta{E}$. For a fixed $\D_t$ and $n_l$, increasing $n_k$ increases the maximum time reached with a fixed number of Trotter steps.
Fig. \ref{fig:egs_kmax}a shows $\D E$ as a function of $n_l$ for different values of $n_k$ and $\D_t{=}0.1$. While for $n_k{=}0$ the value of $n_l$ needs to be higher than $25$ to reach $\D E<10^{-3}$, this value can already be reached with $n_l=7$ for $n_k=3$. This shows that the two-level Trotterization allows to reduce the number of Trotter steps for a given target accuracy, and for the same size of the basis. Importantly, the performance improvement is found for all values of $n_k$ , showing that the method is robust.

We evaluate the behavior for a number of AIM parameters, and find that the two-level multigrid approach always gives smaller $\Delta{E}$ for a given $n_l$. As an illustrative example, in Fig.  \ref{fig:egs_kmax}b we show the result for the AIM parameters at DMFT self-consistency (see Appendix \ref{sec:dmft} for their values). One can see that while the $n_k{=}0$ solution struggles to reach $\D E$ below $10^{-4}$, for $n_k>0$ the error becomes significantly smaller and keeps decreasing for increasing $n_l$. These results confirm that the larger Trotter error of the two-level multigrid approach does not play a significant role in the accuracy of the GS. The reason is that we use these states only as basis states, which does not require that they are very close to the exact time evolution.

\begin{figure}
  \centering
  \includegraphics[scale=0.5]{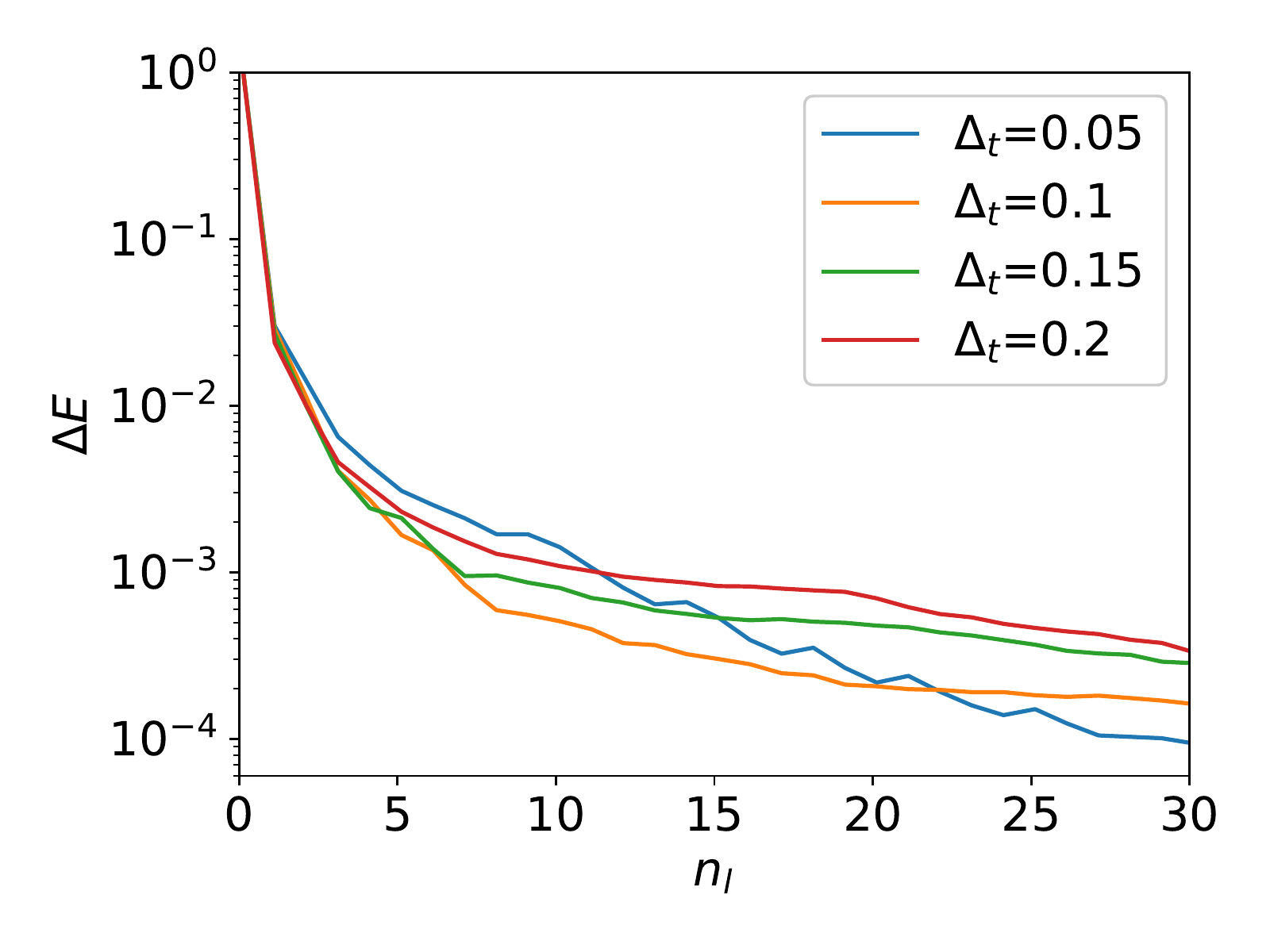}
  \caption{Reduction of $\Delta{E}$ with increasing $n_l$ for different $\D_t$ ($n_k=3$, $U=8$). The AIM parameters are the same as the ones used in Fig. \ref{fig:phi0}}
  \label{fig:egs_dt}
\end{figure}
To further minimize the number of required Trotter steps one may evaluate the behaviour for different values of $\D_t$. Small $\D_t$ gives a more accurate time evolution, but requires more time-steps to reach large times, leading to an increased basis set size. This is reflected in Fig \ref{fig:egs_dt}, which shows that increasing $\D_t$ leads to a progressively larger error for large number of time-steps ($n_l \sim$ 30). On the other hand, a very small $\D_t$ of 0.05 leads to a slower reduction of the error for small $n_l$. For a given $n_l$ of around 10, increasing $\D_t$ therefore has a non-monotonic effect. While increasing $\D_t$ from $0.05$ to $0.1$ improves the result, increasing $\D_t$ further has the opposite effect. The optimal $\D_t$ therefore depends on the value of $n_l$. Overall Fig. \ref{fig:egs_dt} shows that the method is not very sensitive to the detailed choice of $\D_t$, since the general decrease of $\Delta{E}$ with increasing $n_l$ is similar.

\subsection{Green's function}

As mentioned in section \ref{sec:gf_method}, to compute the GF we first need to obtain the ground state wave function and its energy. Based on the GS results of the previous section, we use $\D_t{=}0.1, n_k{=}3, n_l{=}7$ for the first part of the analysis to ensure that $\D E$ is small, since for the considered AIM parameter ranges it gives $\D E$ below $10^{-3}$ (see Fig. \ref{fig:egs_kmax}).
To analyze the quality of the GF, we plot the DOS ($\mathrm{DOS}(\omega)= -\frac{1}{\pi} \text{Im}G(\w + i\d)$) with $\d{=}0.1$.
Since the main ED approximation consists in the discretization of the bath, the peaks in the ED DOS are an approximation for the real DOS, which is usually much smoother. One therefore uses $\d$ to broaden the peaks enough to avoid overly sharp peaks. In terms of accuracy requirements this means that it is typically enough to obtain an energy resolution of similar magnitude as the separation between the individual peaks induced due to the ED discretization.

\begin{figure}
  \centering
  \includegraphics[scale=0.4]{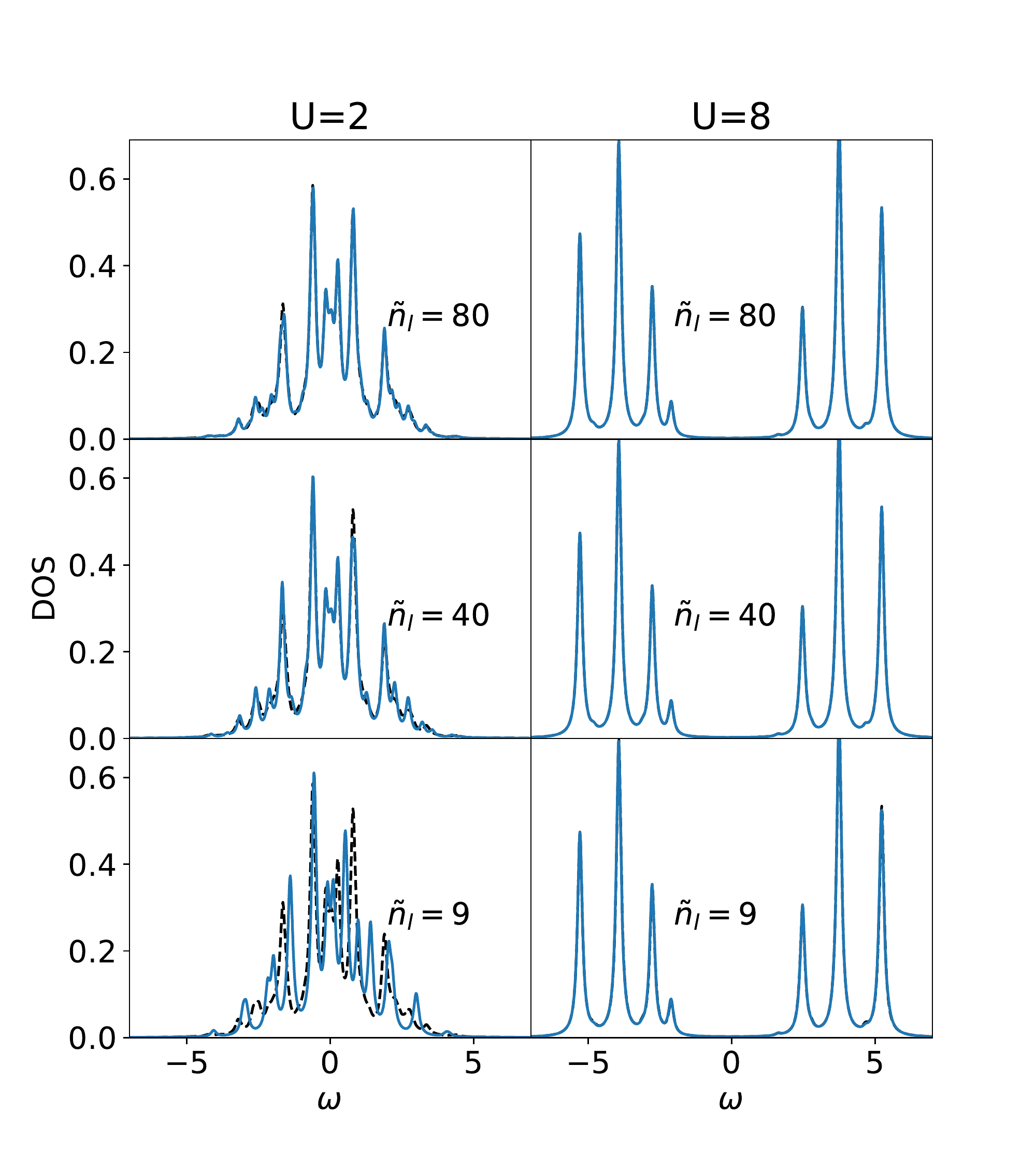}
  \caption{Evolution of the DOS for different $\tilde{n}_l$ used for the GF Krylov basis. The AIM parameters are the ones at DMFT self-consistency (values are given in Appendix \ref{sec:dmft}). The left panels are for $U=2$, giving a metallic state, and the right panels are for $U=8$, giving an insulating state. The solid blue line shows the predicted DOS ($\D_t=\tilde{\D}_t=0.1$, $n_l=7$, $n_k=3$, $\tilde{n}_k=0$), and the dashed black line indicates the exact DOS.}
  \label{fig:dos_lmax}
\end{figure}
In the half-filled case considered here, the Bethe lattice is in a metallic state for $U<6$, while for $U>6$ the Coulomb interaction drives the system to become a Mott insulator \cite{dmft}. We therefore use the values of $U=2$ and $U=8$ to test the QSEG method in both regimes. Fig. \ref{fig:dos_lmax} shows how the DOS evolves for different $\tilde{n}_{l}$ used to construct the GF basis. Here we use a single time-step Trotter expansion ($\tilde{n}_k=0)$. In the left panel we show the metallic regime ($U{=}2$), and in the right panel we show the insulating regime ($U{=}8$). We compare the results to the numerically exact DOS, obtained with a conventional computing ED solver.
For large $\tilde{n}_l{=}80$ there are no visible differences between the exact DOS and the result obtained with QSEG for both the metallic and insulating cases. This shows that the QSEG obtains accurate GFs for both regimes.

\begin{figure}
  \centering
  \includegraphics[scale=0.4]{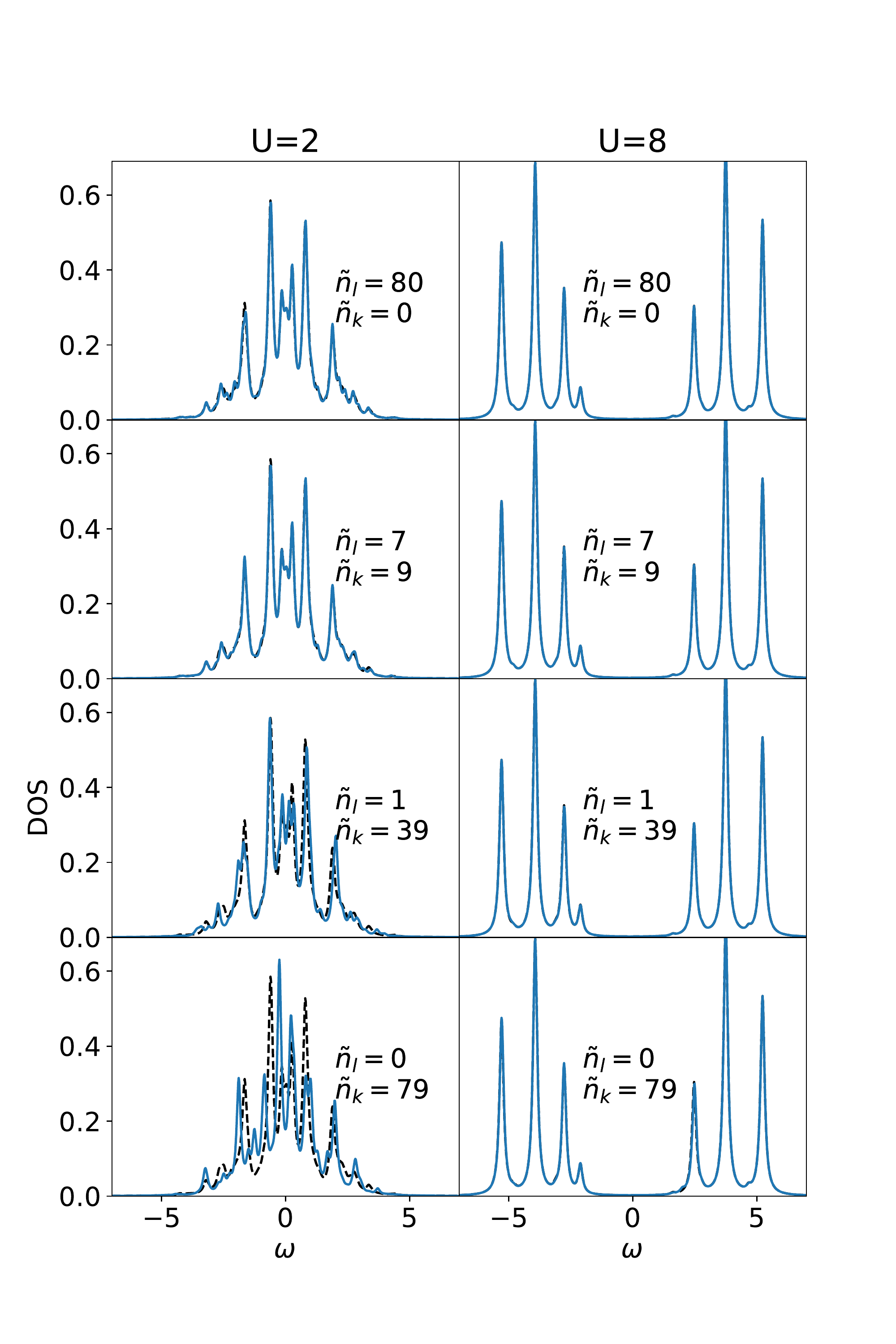}
  \caption{Evolution of the DOS for different $\tilde{n}_l$ and $\tilde{n}_k$ used for the GF Krylov basis, keeping the size of the basis set approximately constant ($n_{\psi} {=} 2(\tilde{n}_l{+}1)(\tilde{n}_k{+}1){-}1$). The AIM parameters are the ones at DMFT self-consistency (values are given in Appendix \ref{sec:dmft}). The left panels are for $U=2$, giving a metallic state, and the right panels are for $U=8$, giving an insulating state. The solid blue line shows the predicted DOS ($\D_t=\tilde{\D}_t=0.1$, $n_l=7$, $n_k=3$), and the dashed black line indicates the exact DOS.}
  \label{fig:dos_kmax}
\end{figure}
As $\tilde{n}_l$ is reduced to $\tilde{n}_l{=}9$ the features of the DOS are still captured qualitatively, but quantitative differences in the metallic case emerge. In the  insulating case there are fewer features, since it is closer to the atomic limit, where the impurity is isolated. In this case, the continued fraction  converges much faster, since the GF has less poles, so that also at low $\tilde{n}_l$ the DOS is already converged.

We now evaluate whether the required number of Trotter steps can be further reduced, while keeping a good accuracy for both the metallic and insulating states, by applying our two-level grid for the Trotter time evolution.
In Fig. \ref{fig:dos_kmax}, the two-level multigrid time evolution basis is used to decrease $\tilde{n}_l$, while keeping the size of the basis in Eq. (\ref{eq:krylov_decomp}), $n_{\psi}{=}2(\tilde{n}_l{+}1)(\tilde{n}_k{+}1){-}1$, approximately constant by increasing $\tilde{n}_k$. We note that, for a given method to prepare the first basis state, the maximum number of Trotter steps to prepare the higher order basis states is determined by the value of $\tilde{n}_l$. For $U{=}8$ a good agreement with the exact solution is found down to $\tilde{n}_l=0$, if the value of $\tilde{n}_k$ is increased accordingly. For $U{=}2$ the computed DOS is in excellent agreement with the exact result down to $\tilde{n}_l=7$ (with $\tilde{n}_k=9$). This is a reduction of the required number of Trotter steps for the GF by a factor of about 9 when compared to the $\tilde{n}_l=80$ required without the two-level grid ($\tilde{n}_k=0$). Also for $\tilde{n}_l{=}1$ the QSEG DOS is still in good agreement with the exact DOS, with some minor quantitative deviations for the metallic state. We note that this requires only $2$ Trotter steps to obtain the Krylov basis for a given initial state. Even for $\tilde{n}_l=0$ the overall agreement is preserved in terms of band widths and band gap, although for the metallic case visible shifts appear in the exact positions and heights of the peaks in the DOS.
For completeness, we also verified that for all bases shown in Fig. \ref{fig:dos_kmax} the full QSEG DMFT self-consistency loop converges to these results.

\begin{figure}
  \centering
  \includegraphics[scale=0.4]{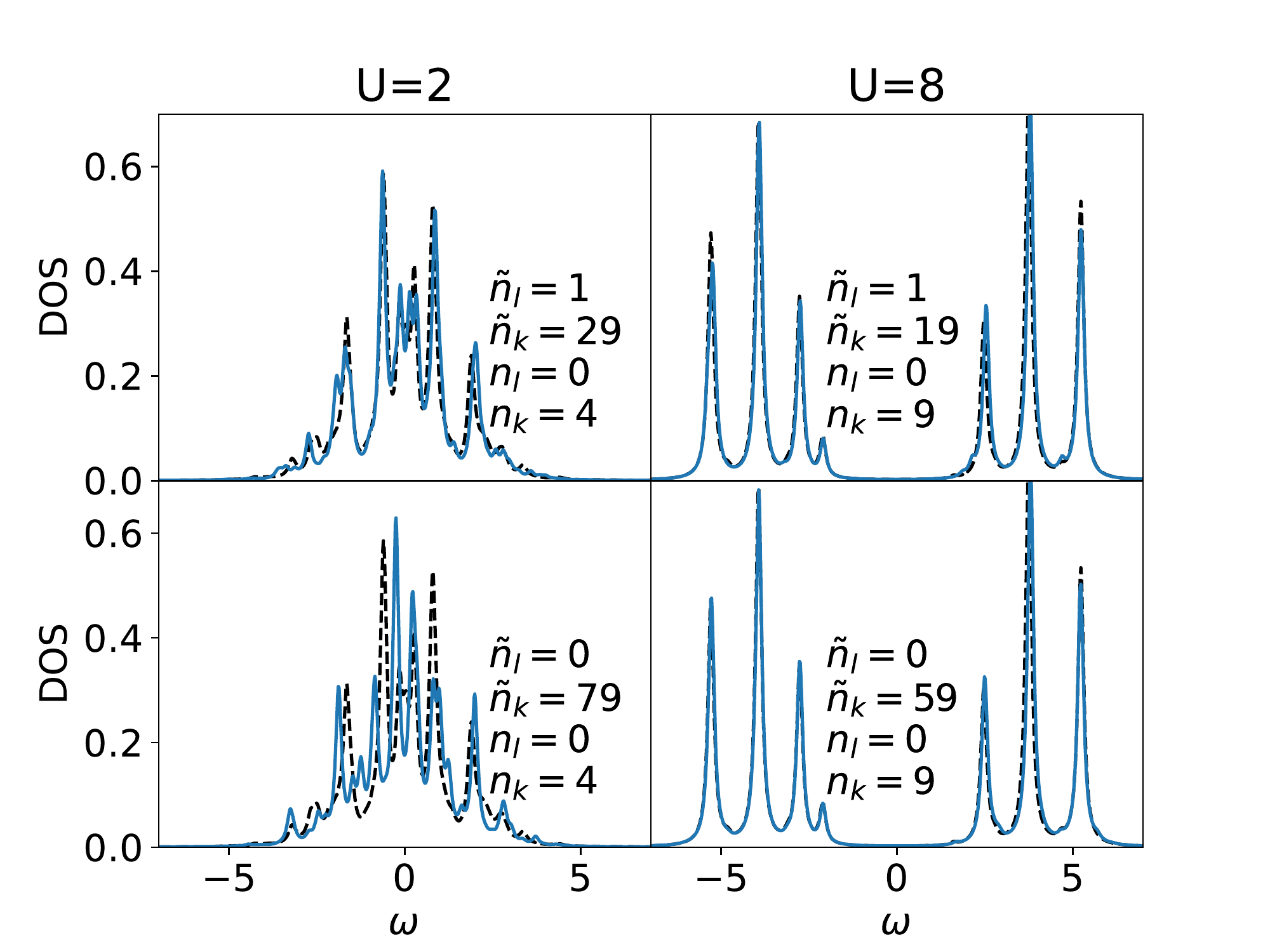}
  \caption{Same as in Fig. \ref{fig:dos_kmax}, but for different QSEG parameters with the lowest possible number of Trotter steps.}
  \label{fig:dos_kmax_differentnlnk}
\end{figure}
Having shown that QSEG can systematically approach the exact solution as $n_l$, $n_k$, $\tilde{n}_l$ and $\tilde{n}_k$ increase, we also evaluate its applicability on near terms devices. In this case one may aim to minimize the number of Trotter steps to be able to cope with the noise in the devices. In the results in Fig. \ref{fig:dos_kmax} we use $n_l{=}7,n_k{=}3$ to ensure that the GS energy error is small for all cases (below $\D E {=}10^{-3}$ eV). Since in QSEG the accuracy of the GF is the main quality parameter, a larger error in the GS is acceptable, provided that the GF is computed to within the desired accuracy. We therefore compute the analogous results to the ones of Fig. \ref{fig:dos_kmax}, but with progressively lower values of $n_l$. We find that for all considered cases even down to $n_l=0$ the DOS is largely identical to the results in Fig. \ref{fig:dos_kmax}. This shows that a single Trotter step is enough to compute the GS wave function to the required accuracy.

In order to minimize the circuit depth one can therefore set $n_l=0$, and then evaluate the quality of the results that can be obtained by the lowest values of $\tilde{n}_l$ of 0 and 1 for different $n_k$ and $\tilde{n}_k$. While increasing $n_k$ and $\tilde{n}_k$ does not increase the circuit depth, it does increase the size of the basis. This leads to improved quality of the results, but also to a larger number of matrix elements that need to be computed. Hence a reduction of the basis set size to a small number for a given accuracy is desirable. We systematically evaluated the required smallest values for $n_k$ and $\tilde{n}_k$ within our two-level multigrid for the Trotter expansion. In Fig. \ref{fig:dos_kmax_differentnlnk} we show examples of parameter settings that allow minimal circuit depths. For $\tilde{n}_l=1$ the DOS is in rather good agreement with the exact values for $\tilde{n}_k{\gtrsim}29$ and  ${n}_k{\gtrsim}4$ for the metallic case, and for $\tilde{n}_k{\gtrsim}19$ and  ${n}_k{\gtrsim}9$ for the insulating case. For $\tilde{n}_l=0$ the insulating state can be recovered well if $\tilde{n}_k$ above 59 is used. For the metallic state, as already discussed earlier, with $\tilde{n}_l=0$ the agreement with the exact solution is reduced.

The number of Trotter steps in the two-level grid to construct a GF basis wave function is equal to $n_l+\tilde{n}_l+2$. The matrix elements in Eq. (\ref{eq:hpsielement}) require the construction of a pair of wave functions, which therefore doubles the total number of Trotter steps. Our results therefore show that one can get accurate results for this system using only 4 Trotter steps for the insulating system ($n_l=\tilde{n}_l=0$), and 6 Trotter steps for the metallic system  ($n_l=0$ and $\tilde{n}_l=1$).
  The corresponding maximum circuit depth is obtained using Eq. \ref{eq:depth}, with $d_V$ and $d_0$ given in Sec. \ref{sec:aim}, and with $N{=}8$, $n_\ua {=} n_{\da}{=} 4$, $n_l{=}0$ and $\tilde{n}_1{=}1$. The resulting maximum number of CNOT layers is 254, and most of the circuits used to evaluate matrix elements will require significantly smaller depths. While this number is still beyond the capabilities of current hardware, we note that the number of used Trotter steps is a few orders of magnitude lower than what needed to obtain the GF by real time evolution and subsequent Fourier transform \cite{troyer}, and hence brings it closer to a practically realizable method as the hardware noise levels are progressively reduced. The significant reduction in the resulting circuit depth requirements is a major advantage of the QSEG method.

\section{Conclusions}
The quantum subspace expansion algorithm for Green’s functions (QSEG) bridges the range of systems between near term and fault tolerant quantum computers, and avoids the quantum circuit parameter optimization necessary for VQAs. It builds on the Lanczos method to compute Green’s functions, which relies on the iterative construction of the Krylov states to obtain the coefficients of the Green’s function in a continued fraction. The Lanczos method is well established on conventional computers, but is limited to small system sizes due to the exponential growth of the Hilbert space with increasing system size. By using a quantum computer to represent the Krylov states, QSEG has the potential to overcome this scaling barrier. QSEG uses the quantum computer to obtain the elements of the Hamiltonian and overlap matrices for the basis states, which can be efficiently prepared on a quantum device by Suzuki-Trotter time evolution. We achieve long time-evolutions with small time steps by using two-level multigrid Trotter time evolution. It takes advantage of the fact that within QSE the time evolution is used to construct a basis, where each basis state does not need to be an accurate time-evolved state. When scaling up to larger systems, the number of matrix elements to compute can become large; hence, it will be important to use efficient representations with minimal basis sizes. The implementation of the method in parallel over many quantum computers will further allow reducing the runtime.

We demonstrate that QSEG can give accurate Green’s functions for the Anderson impurity model on 16 qubits, obtained within DMFT for the Hubbard model on the Bethe lattice with infinite connectivity. The required number of Trotter steps is between 4 for the insulating state and 6 for the metallic state. The QSEG quantum circuit depths are significantly smaller than the ones needed for time-evolution based methods to obtain the Green’s function. This opens up the possibility to run the method on near term devices with low noise levels, while also keeping its potential to scale to large system sizes once fault tolerant quantum computers are available.

\section*{Acknowledgment}

This project was funded and supported by the UK National Quantum Computer Centre [NQCC200921], which is a UKRI Centre and part of the UK National Quantum Technologies Programme (NQTP).

\appendix

\section{Computation of matrix elements of $H$ and $S$}
\label{sec:HS}
In this section, we present how to compute matrix elements  $S_{ij}=\bra{\phi_0}\hat{U}^\dd_i\hat{U}_j\ket{\phi_0}$ and $H_{ij}=\bra{\phi_0}\hat{U}^{\dd}_i\hat{H}\hat{U}_j\ket{\phi_0}$ for the ground state (Eqs. (2) and (3) in the main text), and  $(\bm{S}_{\psi})_{ij}=\bra{\phi_0}\hat{U}^{\dd}_i \hat{\s}_a \hat{U}^{\dd}_j\hat{U}_k \hat{\s}_b\hat{U}_l\ket{\phi_0}$ and $(\bm{H}_{\psi})_{ij}=\bra{\phi_0}\hat{U}^{\dd}_i \hat{\s}_a \hat{U}^{\dd}_j\hat{H}\hat{U}_k \hat{\s}_b\hat{U}_l\ket{\phi_0}$ for the Krylov states (Eqs. (16) and (20) in the main text). Here $\hat{\s}_a$ stands for a Pauli string of the form $\prod_{k=0}^{a-1}\s^{k}_{Z}\s^{a}_{X/Y}$ where $\s^k_{X/Y/Z}$ is a Pauli X, Y or Z operator of qubit k. In general, these matrices may be computed using a Hadamard test \cite{Stair2020,1909.08925}, which requires deep circuits including Toffoli gates.

An alternative approach, which avoids the use of Hadamard tests, has been proposed in Refs. \cite{Kyriienko2020,cortes2021quantum}. The matrix elements of the form $\bra{\phi}\h U_i^{\dd} \h U_j\ket{\phi}$, where $\hat{U}_i, \hat{U}_j$ are unitary operators and $\ket{\phi}$ is a quantum state, can be computed without a Hadamard test by using the so-called multi-fidelity estimation protocol, if one can find a reference state $\ket{R}$ such that:
\begin{itemize}

\item Condition I: $\bra{R}\h U_i^{\dd} \h U_j\ket{R}=r_R e^{i\theta_R}$ is known
\item Condition II: $\bra{R}\h U_i^{\dd} \h U_j\ket{\phi}=0$
\item Condition III: $\ket{\phi}+\ket{R}$ can be constructed.
\end{itemize}
Here $r_R$ is a real number and $\theta_R$ describes the complex phase. Under these conditions we can compute the following state fidelities on a quantum computer:
\begin{align}
  F_1 =&   |(\frac{\bra{\phi} + \bra{R}}{\sqrt{2}})\h U_i^{\dd} \h U_j (\frac{\ket{\phi} + \ket{R}}{\sqrt{2}}) |^2  \nonumber \\=&  \frac{1}{4}|\bra{\phi}\h U_i^{\dd} \h U_j \ket{\phi} +  \bra{R}\h U_i^\dd \h U_j  \ket{R} |^2,\\
F_2=&|\bra{\phi}\h U_i^{\dd} \h U_j \ket{\phi}|^2.
\end{align}
An overview of different methods to compute such state fidelities is presented in Ref. \cite{cortes2021quantum}.
With these equations, one can then obtain $\bra{\phi}\h U_i^{\dd} \h U_j \ket{\phi}=r e^{i\theta}$ as
\begin{align}
  r & = \sqrt{F_2},\\
  \theta & =\arccos{}\Bigg(\frac{4F_1-F_2-r^2_R}{2r_R \sqrt{F_2} }\Bigg) + \theta_R.
\end{align}

We will now outline how the formalism outlined above can be used to compute the matrix elements of interest for the QSEG when the total number of particles is a symmetry of the Hamiltonian.

\subsubsection{Computation of: $S_{ij}=\bra{\phi_0}\hat{U}^\dd_i\hat{U}_j\ket{\phi_0}$}
\label{sec:S}
In the main text we construct the basis for the ground state and the Krylov basis with operators $\hat{U}_i,\hat{U}_j$, which are  Suzuki-Trotter expansions of the Hamiltonian $\h H$, as outlined in Eq. (22) in the main text. If $\h H$ conserves the total number of particles $\hat N$, i.e $[\h H,\h N]=0$,  then $U_i$ and $U_j$ also conserve the number of particles. In this case, condition II can be fulfilled if  $\ket{\phi_0}$ and $\ket{R}$  are eigenvectors of $\h N$ and if they have different number of particles. Here, we choose $\ket{R}$ to be the state with no particles (or all particles in the special case where $\ket{\phi_0}$ contains no particles). With this choice of $\ket{R}$, condition I is also fulfilled,
since  $\hat{U}^\dd_i\hat{U}_j\ket{R}$ has no particles as well, so that it is equal to $\ket{R}$ up to a phase.  This phase is equal to $-h_0 T$, where $T$ is the time used in the Suzuki-Trotter expansion and $h_0$ is the constant term of the Hamiltonian written with fermionic operators. In Eq. 29 in the main text, there is no constant term so the phase is zero.

Finally, condition III depends on the construction of the reference state and is explained in Sec. \ref{sec:HF_prepa} of this document.

\subsubsection{Computation of: $H_{ij}=\bra{\phi_0}\hat{U}^\dd_i\h H \hat{U}_j\ket{\phi_0}$}
\label{sec:H}
For the construction of the matrix elements $H_{ij}$, $\h H$ is decomposed as $\h H {=} \sum_k h_k \h P_k$, where $h_k$ is a coefficient and $P_k$ a Pauli string. Some of the Pauli strings do not conserve the total number of particles, so that $[\h P_k, \h N] \neq 0$. In  the AIM, the terms such as $c_k^\dd c_l+ c_l^\dd c_k=\frac{1}{2} (\hat{\s}^X_k \prod_{k<m<l} \hat{\s}^Z_m\hat{\s}^X_l+\hat{\s}^Y_k\prod_{k<m<l} \hat{\s}^Z_m\hat{\s}^Y_l)$  generate such Pauli strings (for ease of notation, in the rest of the section we drop the Pauli Z strings, which do not change the number of particles). If $\ket{a}$ and $\ket{b}$ are eigenstates of $\h N$, and $|\expval{\h N}{a}-\expval{\h N}{b}|>2$, then $\bra{a}\hat{\s}^{X/Y}_i\hat{\s}^{X/Y}_j\ket{b}=0$, since $\hat{\s}^{X/Y}_i$ connects only adjacent total number of particles sectors.
Therefore, if $\expval{\h N}{\phi_0}>2$, we can choose $\ket{R}=\ket{0} \otimes \ket{0}...\otimes\ket{0}$, while in the special case where if $\expval{\h N}{\phi_0}\leq 2$, we can choose $\ket{R}=\ket{1} \otimes \ket{1}...\otimes\ket{1}$. The motivation behind these choices is to have an $\ket{R}$ state that can be easily obtained in a quantum circuit, and where condition I can be fulfilled.
Finally, condition III depends on the construction of the reference state, which is explained in Sec. \ref{sec:HF_prepa} of this document.

\subsubsection{Computation of: $({S}_{\psi})_{ij}=\bra{\phi_0}\hat{U}^{\dd}_i \hat{\s}_a \hat{U}^{\dd}_j\hat{U}_k \hat{\s}_b\hat{U}_l\ket{\phi_0}$}
\label{sec:Spsi}

For $(\bm{S}_{\psi})_{ij}$, the choice of $\ket{R}$ is the same as in Sec. \ref{sec:H}, since there are 2 Pauli operators which change the total number of particles.

However, more care needs to be taken in condition II, which involves the computation of $\bra{R}\hat{U}^{\dd}_i \hat{\s}_a \hat{U}^{\dd}_j\hat{U}_k \hat{\s}_b\hat{U}_l\ket{R}$. $\hat{U}_l\ket{R}$ is equal to $\ket{R}$ up to a phase $\theta_l$, such that $\bra{R}\hat{U}^{\dd}_i \hat{\s}_a \hat{U}^{\dd}_jU_k \hat{\s}_bU_l\ket{R}{=} e^{i(\theta_l-\theta_i)}\bra{R} \hat{\s}_a \hat{U}^{\dd}_jU_k \hat{\s}_b\ket{R}$. To compute $\bra{\phi_1}\hat{U}^{\dd}_jU_k\ket{\phi_2} $ with $\ket{\phi_1}=\hat{\s}_a\ket{R}$ and $\ket{\phi_2}=\hat{\s}_b\ket{R}$, we can generalize the method presented in Sec. \ref{sec:S} with $\ket{\phi_1}\neq \ket{\phi_2}$. This can be done using the same technique as long as we have $\bra{\phi_1}\ket{R} {=}\bra{\phi_2}\ket{R}{=}0$.

In this case, we have
\begin{align}
  F_1 =&   |(\frac{\bra{\phi_1} + \bra{R}}{\sqrt{2}})\h U_i^{\dd} \h U_j (\frac{\ket{\phi_2} + \ket{R}}{\sqrt{2}}) |^2  \nonumber \\=&  \frac{1}{4}|\bra{\phi_1}\h U_i^{\dd} \h U_j \ket{\phi_2} +  \bra{R}\h U_i^\dd \h U_j  \ket{R} |^2,\\
F_2=&|\bra{\phi_1}\h U_i^{\dd} \h U_j \ket{\phi_2}|^2.
\end{align}

\subsubsection{Computation of: $(H_{\psi})_{ij}=\bra{\phi_0} \hat{U}^{\dd}_i \hat{\s}_a \hat{U}^{\dd}_j\hat{H}\hat{U}_k \hat{\s}_b\hat{U}_l\ket{\phi_0}$}
\label{sec:Hpsi}

The shape of $(\bm{H}_{\psi})_{ij}$ is very similar to the  matrix elements $(\bm{S}_{\psi})_{ij}$, but care needs to be taken for the Pauli string in $\h H$, which may not conserve the symmetry $\h N$. Due to the presence of up to 4 Pauli operators, which connect adjacent symmetry sectors, $\ket{R}$ needs to be chosen such that if $\expval{\h N}{\phi_0}>4$, we choose $\ket{R}=\ket{0} \otimes \ket{0}...\otimes\ket{0}$ and if $\expval{\h N}{\phi_0}\leq 3$, we can choose $\ket{R}=\ket{1} \otimes \ket{1}...\otimes\ket{1}$. We note that it is only valid if the total number of sites is strictly higher than 4.

Finally, condition III depends on the construction of the reference state, and is explained in Sec. \ref{sec:HF_prepa}.

\section{Preparation of the GHZ state with the Hartree-Fock state}
\label{sec:HF_prepa}

As part of condition III, the state $\ket{R} + \ket{\phi_0}$ needs to be implemented, where $\ket{\phi_0}$  is the minimal energy state of $H_0$ with  $n_\ua$ spin-up and $n_\da$ spin-down particles (see Eq. (34) in the main text).
To constuct this superposition, we use a GHZ-state-preparation circuit as suggested in Ref. \cite{cortes2021quantum}. It requires applying a Hadamard gate on the qubit 1 (resp. $N$, where $N$ is the number of sites) and a number $(n_\ua-1)$ (resp. $n_\da-1$) of CNOT gates between qubits $i$ and $i+1$ with $i \in [1,n_\ua]$ (resp. $i \in [N+1,N+n_\da]$). Finally, we apply the $\hat{C}^\dd$ rotation operator in Eq. 34 in the main text to rotate the GHZ state in the local basis. Here,  we used the relation $\hat{C}^\dd \ket{0} {=}\ket{0}$.

The circuit depth in terms of CNOT layers is the sum of the preparation of the layers for the GHZ state preparation, which requires $\mathrm{max}(n_\ua,n_\da)-1$ layers of CNOT gates, and of the $C^\dd$ preparation, which requires $2N$ layers of CNOT gates as explained in Sec. IV of the main text.

\section{ DMFT on the Bethe lattice}
\label{sec:dmft}
For the non interacting solution, the hybridization is given by $\D^0(\w) {=}\frac{1}{2\pi} \Theta (4-\w^2) \sqrt{4-\w^2}$, where $\w$ is the real frequency, and we use the convention $\gamma=1$ as in the main text.

The discretized hybridization is given by $\D^{ED}(z) {=} \sum_{i=n_{\text{imp}}}^{N} \frac{|\e_{0i}|^2}{z - \e_{ii}}$, where we take the hopping matrix to connect a bath site to the impurity as schematically illustrated in Fig. 2 in the main text. The parameters $\e_{ij}$ are chosen to minimize the distance $D = \sum^{n_\mathrm{M}}_{n=0}|\D^{ED}(i\w_n)-\D(i\w_n)|^2$, where the hybridizations are evaluated on the Matsubara axis $\w_n=\frac{(2n+1)\pi}{\b}$, with $\b$ the inverse temperature. Here, we take $\beta{=}100$, which is small enough to keep the zero temperature formalism. Here the role of finite $\beta$ is only to get the hybridization on a finite grid on the imaginary axis, where the function is smoother. In particular, we check that the gap between the first excited state and the  ground state is large enough to have $\e^{\beta (E^{\text{excited}}-E_{\GS})}<0.01$ in order to ignore the population of the excited states compared to the GS. We use $n_\mathrm{M}{=}100$ in order to fit well the low energy part.
Once the Green's function (GF) is determined, we use the DMFT equation $\D^{\text{new}} {=} G$ to find a new bath hybridization, and we reiterate the procedure until the convergence of the GF.

Table \ref{tab:1} (resp. Table \ref{tab:2}) gives the parameters of the first (resp. last) iterations.

\begin{table}[h!]

\begin{tabular}{ccc}
$i$ & $e_{ii}$ & $e_{i1}$\\
2 &-1.17300 &-0.53714\\
3 & -0.37368 &0.38549\\
4 & -0.08996 & -0.21964\\
5  & 0.00000 & 0.13394\\
6  & 0.08996 & -0.21964\\
7  & 0.37368 & 0.38549\\
  8 & 1.17300 & 0.53714\\

\end{tabular}
\caption{\label{tab:1} Bath parameters of the fist DMFT iteration}

\end{table}

Finally, since we are at half-filling, the energy of the impurity site is given by $\e_{11}{=}-\frac{U}{2}$.

\begin{table}[h!]
\begin{tabular}{ccc}
$i$ &$ e_{ii}$ & $e{i1}$\\
2 & -4.78165 & -0.52895\\
3 & -3.09398  & 0.42680\\
4 & -2.17008 & -0.19147\\
5 & -0.00103  & 0.00418\\
6 & 1.61650 & -0.05916\\
7 & 2.74851  & 0.41342\\
  8& 4.63114  & -0.56915\\
\end{tabular}
\caption{\label{tab:2} Bath parameters of the last DMFT iteration}
\end{table}

\newpage
\section{Off-diagonal elements of the Green's function}
\label{sec:offdiagonal}
The algorithm proposed in Sec. IIB of the main manuscript cannot be directly applied to the off-diagonal elements of the greater GF, which are given by
\begin{align}
    G^>_{\a\b}(z)& = \bra{\text{GS}}c_{\a}(z - (H -E_{\text{GS}}))^{-1}c^\dd_{\b}\ket{\text{GS}}.
\end{align}
Here we present the extension of the method to be able to compute these off-diagonal elements. To this aim we first apply the method to obtain the elements of two GFs, defined as
\begin{align}
  G^{(1)}_{\a\b}(z) &= \bra{\GS} (\h c_\a+\h c_\b)(z-\h H+E_\mathrm{GS})^{-1}   (\h c^\dd_\a+\h c^\dd_\b)\ket{\GS},\\
  G^{(2)}_{\a\b}(z) &= \bra{\GS} (\h c_\a-i\h c_\b)(z-\h H+E_\mathrm{GS})^{-1}   (\h c^\dd_\a+i\h c^\dd_\b)\ket{\GS}.
\end{align}
With these modified GFs we then obtain
\begin{align}
  G^>_{\a\b} +G^>_{\b\a} &= G^{(1)}_{\a\b}(z) - G^>_{\a\a}(z)- G^>_{\b\b}(z) \\
  G^>_{\a\b} -G^>_{\b\a} &= i(  G^>_{\a\a}(z)+ G^>_{\b\b}(z) - G^{(2)}_{\a\b}(z))
\end{align}

So $G_{\a\b}^>$ is given by
\begin{equation}
  G_{\a\b}^> = \frac{1}{2}( G^{(1)}_{\a\b}(z) - i G^{(2)}_{\a\b}(z) + (i-1) (G^>_{\a\a}(z)+ G^>_{\b\b}(z)))
\end{equation}

The lesser GF can be computed using the same method.
\bibliography{bibli}

\end{document}